\documentclass[twocolumn,showpacs,preprintnumbers,amsmath,amssymb]{revtex4}

\usepackage{color} 
\usepackage{graphicx}
\usepackage{epsfig}
\preprint{BiMnO3-2015}

\begin{document}
\DeclareGraphicsExtensions{.eps,.jpg,.png}
\input epsf
\title{Optical study of the vibrational and dielectric properties of BiMnO$_3$} 
\author {W. S. Mohamed$^{1}$, A. Nucara$^{2}$, G. Calestani$^{3}$, F. Mezzadri$^{3}$, E. Gilioli$^{4}$, F. Capitani$^{1}$, P. Postorino$^{1}$, and P. Calvani$^{2}$}
\affiliation{$^{1}$Dipartimento di Fisica,  Universit\`{a} di Roma ''La Sapienza'', P.le A. Moro 2, 00185 Roma, Italy}\
\affiliation{$^{2}$CNR-SPIN and Dipartimento di Fisica,  Universit\`{a} di Roma ''La Sapienza'', P.le A. Moro 2, 00185 Roma, Italy}\
\affiliation{$^{3}$Dipartimento di Chimica, Universit\`{a} di Parma, Parco Area delle Scienze 17A, 43124 Parma, Italy}\
\affiliation{$^{4}$CNR-IMEM,  Parco Area delle Scienze 37A, 43124 Parma}

\date{\today}

\begin{abstract}
BiMnO$_3$ (BMO), ferromagnetic (FM) below $T_c \simeq$ 100 K, was believed to be also ferroelectric (FE)  due to a non-centro-symmetric C2 structure, until diffraction  data indicated that its space group is the  centro-symmetric C2/c. Here we present infrared phonon spectra of BMO, taken on a mosaic of single crystals, which are consistent with C2/c at any $T >$ 10 K, as well as room-temperature Raman data which strongly support this conclusion. We also find that the infrared intensity of several phonons increases steadily for $T \to 0$, causing the relative permittivity of BMO to vary from 18.5 at 300 K to 45 at 10 K.  At variance with FE materials of displacive type, no appreciable softening  has been found in the infrared phonons. Both their frequencies and intensities, moreover, appear insensitive to the FM transition at $T_c$.  
\end{abstract}
\maketitle   
   
\section{Introduction}
                                                             
The possible multiferroicity - or, more precisely, magnetoelectricity - of  BiMnO$_3$ (BMO), namely the simultaneous occurrence of ferroelectric (FE) and ferromagnetic (FM) long-range order in this simple perovskite,  has been  long discussed in the literature \cite{Hur,Chapon,Fiebig02,Grizalez,Lotter}. Such interest is justified by its potential applications, which may span from giant electric transformers and multiple-state memory elements, to spintronics, magnetoelectric sensors, electric-field controlled ferromagnetic devices, and variable transducers \cite{Binek,Hill,Eeren,Belik12,Fiebig05,Cheong}.  Indeed, on one hand BiMnO$_3$ is ferromagnetic below \cite{Suga,KimuraPRB} $T_c \simeq$ 100 K, due to superexchange along the Mn$^{3+}$-O$^{2-}$Mn$^{3+}$ chains, with a  maximum  reported magnetization of 3.92 $\mu_B$ per formula unit \cite{KimuraPRB,Calestani07}. On the other hand,  the Bi$^{3+}$ (6s$^2$) lone pair could experience a repulsion from the 2p orbitals of neighboring oxygen ions, leading to a permanent electric dipole and a ferroelectric distortion of the perovskite unit cell. However, both the detection of ferroelectricity, and the observation of such distortion have been long controversial, also for the difficulty to grow single crystals \cite{Toulemonde09} of BMO, being it metastable at ambient pressure \cite{Calestani07}.  
Indeed, the observation of ferroelectricity has been reported  for thin films only \cite{Santos_SSC,Sharan}, where moreover the results strongly depend on their thickness and oxygen stoichiometry \cite{Calestani05}. 

Concerning bulk BMO, it has two structural phase transitions \cite{KimuraPRB}, at 470 K from monoclinic to monoclinic, and at 770 K  from monoclinic to orthorhombic. Both first-principle calculations \cite{Hill} and an experimental study \cite{Santos} supported the FE hypothesis and were consistent with early diffraction studies, which identified a non-centro-symemetric C2 space group symmetry \cite{Atou,Santos}.  However, recent  experimental studies agree that  the symmetry of the whole monoclinic phase at ambient pressure is the centro-symmetric C2/c \cite{Belik07,Calestani07,Toulemonde,Calestani08}. This excludes a ferroelectric phase for BMO, of the displacive type \cite{Scott} at least. However, according to a recent model, inversion symmetry breaking might occur also in the C2/c structure, due to some antiferromagnetic order hidden within canted ferromagnetism \cite{Solovyev}.  Finally, it was found that, for increasing pressure, in single crystals the symmetry evolves to the monoclinic $P2_1/c$ at 1 GPa, and to the orthorhombic $Pnma$ at 6 GPa \cite{Calestani08}.

Valuable information on the crystal symmetry of an insulator can also be provided by  optical studies  (Raman, and/or infrared) of its phonon spectrum. A factor group analysis gives for C2 the representation 29 $A$ + 31 $B$, where 57 phonons are both IR and Raman active while one A and  two B vibrations are acoustic. In the space group C2/c, instead, the representation is 14 $A_g$ + 14 $A_u$ + 16 $B_g$ + 16 $B_u$ where, due to the inversion symmetry,  all the $A_g$ and $B_g$  modes are Raman active, the 13 $A_u$ and 14 $B_u$  modes are infrared active, and the remaining ones are acoustic. Previous infrared measurements of the BMO phonon spectrum \cite{Goian} were fit to the sum of 32  Lorentzian modes. This number is higher than that predicted for the C2/c space group (27), even if still much lower than that expected for the C2 symmetry (57).

\begin{figure}[b]
\begin{center}
\includegraphics[width=8cm]{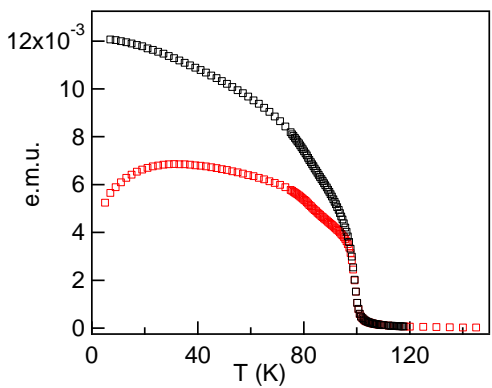}
\caption{Color online. Magnetization vs. temperature (in electromagnetic units, emu) of a single crystal extracted from the mosaic used for the infrared reflectivity measurements. The sample was cooled either in zero field (ZFC) or in a field of 100 Oersted (FC). A sharp and single FM transition occurs at $T_c$ = 100 K.}
\label{Magnet}
\end{center}
\end{figure}

To check and possibly solve this apparent contradiction we have investigated the far-infrared reflectivity of a BiMnO$_3$ mosaic of small single crystals between 10 and 300 K. The resulting optical conductivity could be very well fit at 10 K to a sum of 25 Lorentzians, thus reconciling the results of infrared spectroscopy with a  centro-symmetric structure for BMO. Moreover, we have taken Raman spectra on a single crystal belonging to the same mosaic, finding out that most of the 16 lines observed therein do not coincide with the infrared ones. This provides further support to centro-symmetric structure C2/c, which is also consistent with the observation that no infrared line softens appreciably for $T \to 0$, as it would occur in displacive ferroelectrivity. Finally, we observe a remarkable increase in the infrared oscillator strength for lowering temperature,   which causes the extrapolated relative premittivity to increase smoothly  from 18.5 at 300 K to 45 at 10 K.   

\begin{figure}[b]
\begin{center}
{\hbox{\epsfig{figure=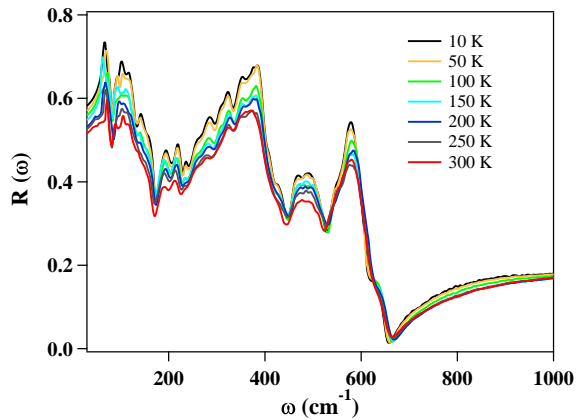,width=8cm}}}
\caption{Color online. Reflectivity spectra of bulk BiMnO$_3$, in the far-infrared range of frequencies at different temperatures.}
\label{R}
\end{center}
\end{figure}

 \section{Experiment and results}
 
The BiMnO$_3$ sample was grown and characterized as described in Ref. \onlinecite{Calestani08}. The compound was  synthesized by solid state reaction of Bi$_2$O$_3$ (Aldrich 99.99 \%) and Mn$_2$O$_3$ (Aldrich 99.999 \%), with the latter one in slight excess,  at 4 GPa and 1073 K, using a high-pressure multi-anvil apparatus to stabilize the phase. We thus obtained a mosaic, whose dimensions are about 5 $\times$ 4 $\times$ 1.5 mm, made of  single crystals too small to be used individually for the far-infrared measurements. Some of them were then extracted to perform the structural and magnetic measurements, while the rest of the batch was polished with powders having grain size down to 0.5 $\mu$m for the infrared experiment. Typical magnetization curves in field cooling (FC, 100 Oe) and zero-field cooling (ZFC)  are shown in Fig. \ref{Magnet}. They display a single and sharp FM transition at 100 K, which demonstrates the good chemical quality of the sample.

\begin{figure}[b]
\begin{center}
{\hbox{\epsfig{figure=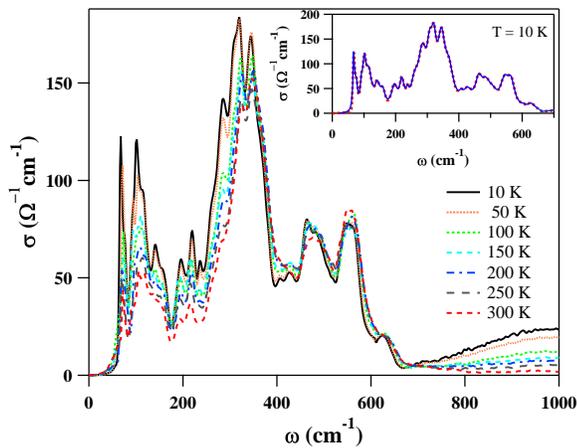,width=8cm}}}
\caption{Color online. Optical conductivity of bulk BiMnO$_3$ in the far infrared range, at different temperatures. The inset shows the fit (dotted line) to the optical conductivity data (solid line) used to obtain the parameters at 10 K listed in Table I.}
\label{sigma}
\end{center}
\end{figure}

The sample reflectivity $R(\omega)$ was measured by an interferometer Bruker 66V, coupled  to an Infrared Labs. liquid-He cooled bolometer,  a Hg-Cd-Te detector, or a Si photodiode, depending on the spectral range, with a spectral resolution of 2 cm$^{-1}$ in the phonon region. The radiation was unpolarized: even if the whole sample were a single crystal, the monoclinic structure of BMO would  prevent to obtain more meaningful information by polarized radiation.  The sample temperature was regulated between 10 and 300 K with a thermal stability and accuracy of $\pm 2$ K. The reference was a golden mirror  close to the sample and oriented by a He-Ne laser parallel to it. Even if the  region of interest  is limited here to 50 $\div$ 700 cm$^{-1}$, the spectral range was extended from 30 to 10000 cm$^{-1}$ in order to obtain accurate Kramers-Kronig (KK) transformations \cite{Dressel}.The extrapolation from $\omega$ = 30  cm$^{-1}$ to $\omega$ = 0 was performed by a constant $R(\omega)$, as usual for good insulators, that to $\omega = \infty$ by a $\omega^{-4}$ power law \cite{Yang}. 
The Raman spectra were measured with a  Horiba LabRAM HR Evolution  micro-spectrometer in  backscattering geometry. The samples were different small crystals belonging to an unpolished area of the same mosaic where the infrared spectra were measured. They were excited by the 632.8 nm radiation of a He-Ne laser with 30 mW output power. Also the Raman spectra, as the infrared ones,  were taken in unpolarized configuration.  The detector was a Peltier-cooled Charge-Coupled Device (CCD)  and the resolution 1 cm$^{-1}$, thanks to a 1800 grooves/mm grating with 800 mm focal length. Measurements were performed with a 20x objective (numerical aperture NA=0.35).

\begin{table} 
\caption{Infrared phonon frequencies $\Omega_{j}$,  oscillator strengths $S_j$, and widths $\Gamma_{j}$, as obtained by fitting to Eq. (1) the experimental $\sigma(\omega)$  at 10 K and 300 K. All frequencies and widths are in cm$^{-1}$, intensities in cm$^{-2}$.}

\begin{ruledtabular}
\begin{tabular}{ccccccc}

$j$ & $\Omega_{j}$(10 K) &  $S_{j}$(10 K) & $\Gamma_j$(10 K) & $\Omega_{j}$(300 K) & $S_{j}$(300 K) & $\Gamma_j$(300 K) \\
\colrule

1      & 68		& 39000        & 6	& 72	& 19000        &  8  \\
2	& 75		& 30000        & 10	&	&	           &      \\			
3	& 92		& 21000	& 9	& 92	& 20000	& 16  \\
4	& 101	& 60000	& 12	& 105	& 16000	& 11 \\
5	& 115	& 100000	& 23	& 117	& 50000	& 24 \\
6	& 143	& 54000	& 21	& 139	& 44000	& 31 \\
7	& 159 	& 26000	&16	& 157	& 19000	& 20 \\
8	& 196	& 57000	& 22	& 192	& 17000	& 18 \\
9	& 219	& 49000	& 17	& 216	& 32000	& 24 \\
10	& 237	& 12000	&10	& 235	& 3000	& 12 \\
11	& 261	& 78000	& 29	& 260	& 39000	&32 \\
12	& 284	& 178000	& 30	& 280	& 42000	& 31 \\
13	& 308	& 115000	& 22	& 304	& 76000	& 38 \\
14	& 321	& 90000	&18	& 321	& 80000	& 24 \\
15	& 345	& 230000	& 30	& 346	& 120000	& 50 \\
16	& 369	& 110000	& 31	& 371	& 100000	& 44 \\
17	& 407	& 7500	& 12     &         &		&      \\		
18	& 426	& 56000	& 34	& 421	& 40000	& 46 \\
19	& 463	& 80000	& 28	& 462 	&39000	& 30 \\
20	& 486	& 110000	& 35	& 489	&60000	& 45 \\
21	& 508	& 48000	& 38	&	& 	           &      \\
22	& 541	& 47000	& 22	& 545 & 46000        &  25 \\		
23	& 561	&125000	& 35	& 565	& 122000	& 39 \\
24	& 604	& 3000	& 46   &         &		&      \\			
25	& 628	& 18000	& 26	& 629	& 10000	& 28 \\

\label{Table I}
\end{tabular}
\end{ruledtabular}
\end{table} 


The reflectivity of BiMnO$_3$ is shown in Fig. \ref{R} in the far infrared range at all temperatures and is basically similar to that reported in Ref. \onlinecite{Goian}, including a strong increase of the intensity of some lines for $T \to$ 0.  Above that range, R($\omega$) looks flat and basically independent of temperature within the errors. 
                                          
The optical conductivity  $\sigma(\omega)$ extracted by the KK transformations from $R(\omega)$ is shown in Fig. \ref{sigma}.   The phonon parameters were obtained by fitting  $\sigma(\omega)$  to the sum of Lorentzian oscillators 

\begin{equation}
\sigma_1 (\omega) = \frac{1}{60}\sum_{j=1}^{n}\frac{\omega^2 \Gamma_jS_{j}}{(\Omega^{2}_j-\omega^{2})^2+\omega^{2}\Gamma_i^2} \\
\label{sigma_fit}
\end{equation}
\

\noindent
where the factor 1/60 allows for $\sigma_1 (\omega) $ being measured \cite{Timusk} in  $\Omega^{-1}$ cm$^{-1}$. Here, $\Omega_j$ and $\Gamma_j$ 
are the peak frequency and the linewidth of the $j$-th transverse optical mode,  respectively, in cm$^{-1}$, and  $S_j$ is  the oscillator strength in cm$^{-2}$. The fitting parameters are listed in Table I for both the highest and the lowest temperature. Excellent fits (like that the inset of Fig. \ref{sigma}) were obtained at 10, 100, and 150 K for $n$ = 25 oscillators, confirming the diffraction result that no structural change is associated with the FM transition. At 300 K, due to line broadening which superimposes some modes, only 21 oscillators were necessary. As already mentioned, in Ref. \onlinecite{Goian}, the infrared phonon spectrum was fit using instead $n$  = 32 modes, a value  not compatible with the centro-symmetric C2/c. This discrepancy was tentatively explained by the authors in terms of overtones or combination bands - which however should have intensities much lower than the main lines, for the usual values of the anharmonic potential. The fact that here we get accurate fits  down to 30 cm$^{-1}$ for $n$ = 25, a number of modes smaller than the 27  predicted for the centro-symmetric  structure C2/c, makes our results  consistent with those  of diffraction  \cite{Belik07,Calestani07,Toulemonde}.

\begin{figure}[b]
\begin{center}
{\hbox{\epsfig{figure=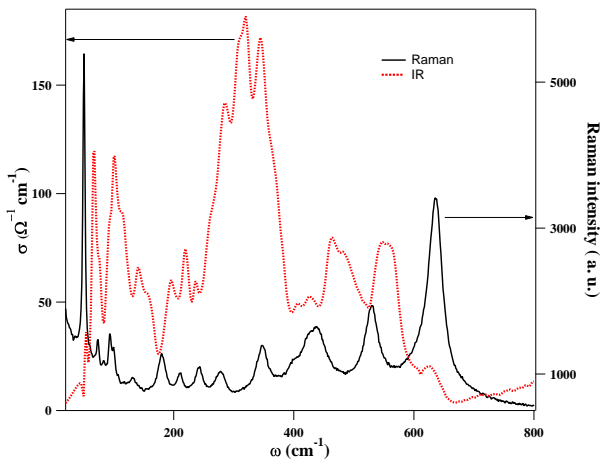,width=8cm}}}
\caption{Color online. Raman spectrum of a single crystal of BiMnO$_3$ at room temperature and infrared spectrum of the mosaic at 10 K, extracted from Fig. \ref{sigma}. Given the weak $T$-dependence observed in both spectra (see Table I and Ref. \onlinecite{Toulemonde}) the best resolved IR spectrum has been chosen for a more meningful comparison.}
\label{raman}
\end{center}
\end{figure}

According to well known selection rules,  the modes which are infrared-active in a centro-symmetric structure are not Raman active, and vice versa. Therefore, to further check the BMO cell symmetry, we examined the Raman spectra of three micro-crystals, which gave the same results. The experiment was done at room temperature, considering that diffraction data exclude phase transitions below 300 K \cite{Calestani08}, and that neither the Raman lines in Ref. \onlinecite{Toulemonde}, nor the  infrared modes in Table I display major frequency shifts down to the lowest temperatures. A typical Raman spectrum is shown in Fig. \ref{raman}, where sixteen lines are observed, out of the predicted thirty.  Three out of them, at or above 400 cm$^{-1}$, are poorly resolved. The data are very similar to those  reported for room temperature in Ref. \onlinecite{Toulemonde}, but thanks to a somewhat larger experimental range we observe an additional sharp line at 51 cm$^{-1}$. In the same Fig. \ref{raman} we  report  also the best resolved infrared spectrum of the mosaic in Fig. \ref{sigma},  that at 10 K. The two spectra look quite different and, except for a few features where the finite linewidths do not allow for a certain claim, the Raman lines do not coincide with the infrared ones. The Raman phonon frequencies  and widths, obtained by  fitting to data a sum of Lorentzians, are listed in Table II, which allows also for a comparison  with the IR frequencies in Table I at 300 K. These results provide further support to our previous conclusion that the ions of BMO  vibrate in  a centro-symmetric cell. 

Another interesting feature of the spectra in Fig. \ref{sigma} is their temperature dependence. The frequencies do not change appreciably with $T$, and no phonon softening is observed which may indicate displacive ferroelectricity - not even incipient as in SrTiO$_3$ \cite{Yamanaka}. 
Moreover, we do not observe any appreciable shift  in the $\omega_j$'s around or below the FM transition. In Ref. \onlinecite{Toulemonde}, a low-frequency shift was reported for the Raman bands at 513.8 cm$^{-1}$ and 637.4 cm$^{-1}$ below $T_c$. The authors attribute this observation to a spin-phonon coupling which is switched on by the FM transition. The absence of a corresponding effect in the present infrared data indicates that such interaction, if may be able to change the Raman polarizability, does not affect appreciably the first-order interaction of the radiation with the lattice dipole moment.

\begin{figure}[b]
\begin{center}
{\hbox{\epsfig{figure=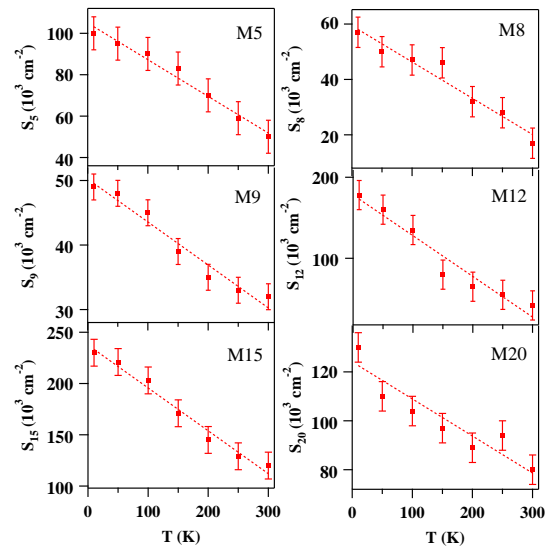,width=8cm}}}
\caption{Color online. Systematic increase of the oscillator strength of BiMnO$_3$ phonons  for $T \to 0$. The modes (M) are numbered as in Table I and the lines are guides to the eye. Within errors, no reproducible features are observed around the FM transition at $T_c \simeq$ 100K.}
\label{S}
\end{center}
\end{figure}

However, as we already mentioned and is evident in Table I, the oscillator strengths $S_j$ of some lines (with $j$ = 1, 4, 5, 8, 11, 12, 15, 20, 25)  increase by about a factor of two, or even more, from 300 to 10 K, an effect that can be observed also in Fig. 2 of Ref. \onlinecite{Goian}.  Fits to our data at all temperatures show that  the intensity of those phonons linearly increases for decreasing temperature, as shown in Fig. \ref{S}, and is insensitive to the  FM transition at $T_c$. Similar effects have been observed in several insulating oxides with high polarizability \cite{Homes2001,Homes2003,Calvani95,BCGO14} and often attributed  to the so called ``charged phonon'' effect \cite{Rice1979} by which phonons acquire an anomalously large spectral weight due to a redistribution of electronic charge, within the cell, associated with their displacement pattern. Such increase of spectral weight in the far-infrared region causes the failure of the $f$-sum rule 

\begin{equation}
\int_0^{\omega_{max}}{\sigma(\omega,T)d\omega} = const.
\label{sum}
\end{equation}

\noindent
if $\omega_{max}$ does not include the regions where the transferred spectral weight comes from, presumably the electronic bands in the visible and the UV, which are out of the present measuring range. Anyway, checking such assumption  would be a very hard task, given the small width of the FIR range (where the spectral weight in Eq. \ref{sum} increases for lowering temperature) with respect of the spectral regions where it should decrease. The same holds for the weak band partially shown in Fig. \ref{sigma},  at frequencies higher than the phonon range. It can either result from an admixture of phonon overtones and combination bands, consistently with its low  intensity, or have a polaronic origin, namely, to come from excess charges self-trapped in the lattice. Those charges could be related to oxygen non-stoichiometry, which is likely to occur in this oxide \cite{Solovyev2}. In both cases the band intensity will increase for T $\to$ 0: in the former one because it will follow the  strengthening of the individual phonons, in the latter one because, at low $T$, more  quasi-free charges will self-trap in the local distortions \cite{Millis,Calvani}.

\begin{table} 
\caption{Raman phonon frequencies $\Omega_{j}$ obtained by fitting to Eq. (1) the BiMnO$_3$ Raman intensity  at 300 K. All frequencies and widths are in cm$^{-1}$.}

\begin{ruledtabular}
\begin{tabular}{ccc}

Phonon ($j$) & $\Omega_{j}$ (300 K) & $\Gamma_j$ (300 K) \\
\colrule

1      & 50		& 2	  \\
2	& 74		& 1	  \\			
3	& 84		&  1	  \\
4	& 93 	& 1 	  \\
5	& 99		& 1 	 \\
6	& 133	& 6	 \\
7	& 180 	& 9	 \\
8	& 211	& 9	 \\
9	& 242	& 10	 \\
10	& 278	& 15	 \\
11	& 347	& 15	 \\
12	& 400	& 15	 \\
13	& 424	& 12	 \\
14	& 440	& 17	 \\
15	& 530	& 12	 \\
16	& 635	& 14	 \\

\label{Table I}
\end{tabular}
\end{ruledtabular}
\end{table} 


\begin{figure}[b]
\begin{center}
{\hbox{\epsfig{figure=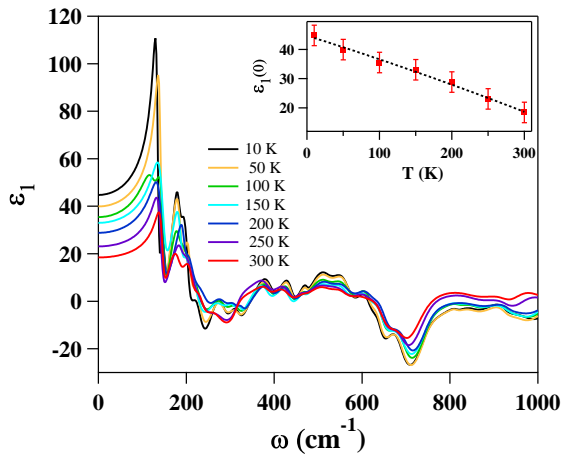,width=8cm}}}
\caption{Color online. Behavior with frequency and temperature of the real part of the dielectric function $\epsilon_1(\omega)$  in BiMnO$_3$. The inset shows the temperature dependence of its extrapolation to zero frequency, the (relative) dielectric constant  $\epsilon_1(0)$.}
\label{Eps1}
\end{center}
\end{figure}

Together with the optical conductivity, which is related to the imaginary part of the dielectric function $\tilde \epsilon(\omega)$, the  KK transformations  provide its real part $\epsilon_1(\omega)$, that is shown in Fig. \ref{Eps1}. If one assumes the absence of strong modes at  frequencies lower than the limit of this experiment (30 cm$^{-1}$), its extrapolation to zero frequency gives the relative permittivity $\epsilon_1(0)$, which can be thus be determined vs. temperature without any metal deposition and contacts on the sample, which may affect the measurement. As shown in the inset of the same Figure, due to the above described growth of several phonon intensities at low $T$, it increases smoothly from 18.5 at 300 K to 45 at 10 K. These rather low values support once again the absence of a ferroelectric phase in BiMnO$_3$ down to 10 K.

 \section{Conclusion}
 
In order to further investigate the possible occurrance of ferroelectricity in BiMnO$_3$, we have measured the infrared phonon spectrum of a mosaic of small single crystals from 300 to 10 K. We find that at the lowest $T$ it can be accurately fit by using only 25 Lorentz oscillators, out of the 27 predicted for the C2/c symmetry that has been attributed to BiMnO$_3$ by the most recent diffraction experiments. To further check the conclusion that the BMO cell is centro-symmetric we have  measured the Raman spectrum at 300 K of three single crystals belonging to the mosaic, and we have observed sixteen phonon lines out of the thirty predicted for C2/c. Most of them are definitely different from the infrared lines, while for a few others, the finite linewidths in both spectra do not allow for an unambiguous conclusion. This is consistent with a centro-symmetric structure like C2/c while, in case of broken inversion symmetry, the Raman and infrared combs should substantially coincide. If one adds that our data do not show any appreciable phonon softening, the  present results  point consistently toward the absence of ferroelectricity - and therefore of multiferroicity - in BiMnO$_3$ down to 10 K, at least of displacive type. 

However, we observe a regular increase of the oscillator strength in several modes of  BiMnO$_3$ for lowering temperature. This causes the extrapolation to zero frequency of the real part of the dielectric function, namely, the relative permittivity, to increase from 18.5 at 300 K to 45 at 10 K.  Across the ferromagnetic transition at $T_c \simeq$ 100 K we do not detect any major anomaly in the oscillator strength, nor in the phonon frequency. Therefore, we conclude that the phonon-spin interaction recently invoked to interpret the behavior of a few Raman phonon lines does not affect the dipole moment which interacts at first order with the radiation field.

\end{document}